\documentclass[nofootinbib,aps,a4paper,10pt,superscriptaddress,twocolumn,showkeys,eqsecnum]{revtex4-1}
\usepackage{amsmath}
\usepackage{amsmath,tabstackengine}
\usepackage{amsfonts}
\usepackage{tabularx}
\usepackage{framed}
\usepackage{amsbsy}
\usepackage{enumitem}
\usepackage{times}
\usepackage{colortbl}
\usepackage{array}       
\usepackage{tabularx}     
\usepackage{lipsum} 
\usepackage{array}
\usepackage{multirow}
\usepackage{makecell}
\usepackage{booktabs}
\usepackage{graphicx}
\usepackage{multirow}
\usepackage{color}
\usepackage{braket}
\usepackage{dcolumn}
\usepackage{siunitx}
\usepackage{bm,url}
\usepackage{colortbl}
\usepackage[usenames,dvipsnames,svgnames]{xcolor}  %% colored text
\usepackage{hyperref}   %% Needs to make hyper reference of any references or citations
\definecolor{oxfordblue}{rgb}{0.0, 0.13, 0.28}
\definecolor{burgundy}{rgb}{0.5, 0.0, 0.13}
\definecolor{darkolivegreen}{rgb}{0.33, 0.42, 0.18}
\definecolor{darkblue}{rgb}{0,0,0.5}
\definecolor{richcarmine}{rgb}{0.84, 0.0, 0.25}
\definecolor{darkblue}{rgb}{0,0,0.5}
\definecolor{bluer}{rgb}{0.00,0.50,0.75}{}
\hypersetup{colorlinks=true, citecolor=red, linkcolor=blue,
	urlcolor = magenta, filecolor=magenta}
\begin{document}
	% Title and author information
	\newcommand\be{\begin{equation}}
		\newcommand\ee{\end{equation}}
	\newcommand\bea{\begin{eqnarray}}
		\newcommand\eea{\end{eqnarray}}
	\newcommand\bseq{\begin{subequations}} %solo con amsmath
		\newcommand\eseq{\end{subequations}}
	\newcommand\bcas{\begin{cases}}
		\newcommand\ecas{\end{cases}}
	\newcommand{\p}{\partial}
	\newcommand{\f}{\frac}
	% Define parameters
		% Parameters - use values that keep the vector field bounded

	\title{A winding number analysis of Schwarzschild black hole stability in light of Planck-scale modified kinematics }
	\author{\textbf{Mohsen Khodadi}}
	\email{m.khodadi@du.ac.ir}
	\affiliation{School of Physics, Institute for Research in Fundamental Sciences (IPM),	P. O. Box 19395-5531, Tehran, Iran}
	\affiliation{School of Physics, Damghan University, Damghan 3671641167, 
		Iran}
	\affiliation{Center for Theoretical Physics, Khazar University, 41 Mehseti Str., AZ1096 Baku, Azerbaijan}
\author{\textbf{Nosratolla Jafari}}
\email{nosrat.jafari@fai.kz}
\affiliation{Fesenkov Astrophysical Institute, 050020, Almaty, Kazakhstan}
\affiliation{Al-Farabi Kazakh National University, 050040 Almaty, Kazakhstan}
	
\author{\textbf{Shahin Mamedov}}
\email{	ctp@khazar.org}
\affiliation{Center for Theoretical Physics, Khazar University, 41 Mehseti Str., AZ1096 Baku, Azerbaijan}
\affiliation{Institute for Physical Problems, Baku State University, Z.Khalilov 23, Baku, AZ-1148, Azerbaijan}
\affiliation{
	Institute of Physics, Ministry of Science and Education, H.Javid 33, Baku, AZ-1143, Azerbaijan}

\begin{abstract}
Determining whether Planck-scale effects can stabilize black holes addresses fundamental questions about black hole evaporation and quantum gravity consistency. Here, we analyze the thermodynamic topology of Schwarzschild black holes under Planck-scale modified kinematics, using a cubic entropy correction derived from a well-known phenomenological MDR with leading correction \(\eta E^3/E_P\). Enforcing physical constraints (\(S'(r_h) > 0\), \(T > 0\)) via the entropy-geometry correspondence, we find a single unstable branch with \(w = -1\) and \(W = -1\) for both signs of the correction parameter. A second root suggesting stability (\(w = +1\)) is excluded due to negative mass/temperature and lies outside the perturbative regime. Thus, this class of MDRs does not yield stable Schwarzschild black holes. However, MDRs with different leading-order corrections may behave otherwise, leaving the search for Planck-scale stabilization an open endeavor.

\vspace{0.5cm}

\textbf{Keywords:} Winding numbers, Phase stability, Schwarzschild black hole, Emergent-gravity paradigm
\end{abstract}
	\maketitle
\section{Introduction}
\label{sec:intro}

Numerous theoretical approaches have been proposed to resolve the fundamental incompatibilities between General Relativity (GR) and Quantum Field Theory (QFT). Notable examples include string theory \cite{Gross:1987ar,Amati:1988tn,Mukhi:2011zz}, loop quantum gravity \cite{Rovelli:1994ge,Ashtekar:1996eg}, and causal set theory \cite{Wallden:2013kka}. A recurring feature in many of these frameworks is the introduction of a minimum measurable length, typically associated with the Planck scale, which correspondingly implies a maximum energy scale (the Planck energy) \cite{Garay:1994en,Calmet:2004mp}. A theory of quantum gravity (QG) must replace classical spacetime with a quantum structure. One key characterization of this structure is the existence of a fundamental minimum length \footnote{A robust area of research investigates the theoretical and phenomenological consequences of the minimum-length scale that emerges as a natural ultraviolet cutoff in QG models (e.g., \cite{Ali:2011ap,FaragAli:2015boi,Nozari:2015qoi,Khodadi:2017eim,Khodadi:2016gyw,Khodadi:2018scn,Tee:2022xgm,Casadio:2022opg}).}, implying that at the smallest scales, spacetime loses its definite shape. This quantum nature also suggests that Lorentz invariance—a cornerstone of classical relativity—must be broken or deformed at high energies. Such modifications could lead to novel phenomena like the spontaneous formation of micro black holes. Furthermore, spacetime at this scale is theorized to be a dynamic foam, where virtual particle-antiparticle pairs constantly fluctuate into and out of existence \cite{Ng:2011rn}. These profound changes could manifest as modified particle dynamics or the emergence of nonlocal interactions.

Several avenues have been explored in the pursuit of a consistent quantum gravity theory. A foundational approach modifies GR by extending the Einstein-Hilbert action with functions of the Ricci scalar, known as $f(R)$ gravity \cite{Sotiriou:2008rp,Nojiri:2010wj}. While phenomenologically rich, these theories generally fail to be renormalizable \footnote{Progress toward renormalizability is made by incorporating terms quadratic in the curvature tensors (Ricci and Riemann) \cite{Salvio:2018crh}. Yet, this improvement comes at the cost of a Hamiltonian that is not bounded from below, leading to fatal physical pathologies like vacuum instability and non-unitarity. A promising resolution to this trade-off is found in theories that include infinite derivatives of curvature invariants, thereby defining the framework of infinite derivative gravity \cite{Biswas:2011ar}.}. Departing from curvature-based modifications, a distinct class of theories introduces a high-energy scale $\Lambda$ to alter the foundational kinematics of Special Relativity (SR). This leads to two broad phenomenological frameworks defined by their treatment of spacetime symmetries: Lorentz Invariance Violation (LIV), and Doubly Special Relativity (DSR). In the former, Lorentz symmetry is considered an emergent low-energy phenomenon. At scales near $\Lambda$, a preferred frame appears, violating the relativity principle. The dispersion relation is modified, but momentum remains additively conserved \cite{Colladay:1998fq}. On the latter, the relativity principle is preserved, but its realization is deformed at high energies \cite{Amelino-Camelia:2000bxx}. The core of DSR \cite{Amelino-Camelia:2000cpa,Amelino-Camelia:2000stu,Amelino-Camelia:2003cem,Kowalski-Glikman:2001vvk,Amelino-Camelia:2011lvm} is a triad of compatible modifications: a set of Modified Dispersion Relations (MDRs), a nonlinear momentum composition law—known as the Modified Composition Law (MCL)—and a set of deformed Lorentz transformations that leave the first two ingredients invariant for all inertial observers.

More exactly, DSR is a class of phenomenological theories—a bottom-up approach to QG—that extend SR by introducing an additional invariant scale beyond the speed of light, typically the Planck energy, while retaining the relativity principle \cite{Magueijo:2001cr}. The original motivation stems from attempts to reconcile the seemingly universal nature of the speed of light with the appearance of a fundamental length (or energy) scale in QG. Unlike LIV frameworks, DSR preserves the relativity of inertial frames but modifies the transformation laws in such a way that the Planck energy becomes an invariant maximum energy for particles, much like $c$ is an invariant maximum speed.

Within both LIV and DSR, the kinematics are typically modified via a general function of the dimensionless ratio \(E/E_{\text{Pl}}\), leading to a dispersion relation of the form \(E^2\left(1 + f(E/E_{\text{Pl}})\right) = p^2 c^2 + m^2 c^4\) \cite{Amelino-Camelia:1997ieq} \footnote{It is important to note that DSR encompasses a family of distinct realizations, each characterized by a specific form of the dimensionless function $f(E/E_{\text{Pl}})$ in the MDR \cite{Jafari:2020ywd}.}. The critical distinction is that such an MDR, taken in isolation, is phenomenologically degenerate between the two frameworks. To break this degeneracy, one must additionally specify: (i) the law of momentum composition—whether it remains linear (LIV) or becomes nonlinear (DSR)—and (ii) how the dispersion relation transforms between observers—i.e., whether it is frame-dependent (LIV) or remains invariant under deformed Lorentz transformations (DSR)—for more details see Refs. \cite{Myers:2003fd,Kowalski-Glikman:2002eyl,Amelino-Camelia:2002cqb,Magueijo:2002am}.

Mathematically, DSR finds a natural formulation in the language of Hopf algebras and non-commutative geometry \cite{Majid:1994cy}. In this setting, the MCL corresponds to an algebraic coproduct \cite{Majid:1994cy,Carmona:2016obd}, as exemplified by the $\kappa$-Poincar\'e model \cite{Majid:1999tc}, where the deformation parameter $\kappa$ is identified with the Planck scale and the composition law is noncommutative yet associative, allowing a well-defined multiparticle extension via pairwise composition. Phenomenologically, DSR effects are expected to become significant near the Planck scale, but they can leave imprints in a variety of physical systems (e.g., see \cite{Guvendi:2025det,Jafari:2025qgc,Jafari:2025rar,Boumali:2025bpk,Khodadi:2019hsy}).

%\paragraph{Scope of the present work.}
In this paper, we do \emph{not} commit to a specific DSR model nor attempt to break the LIV-DSR degeneracy. Instead, we adopt a phenomenological MDR of the form (\ref{eq:dsr_dispersion}) as a common low-energy parametrization of Planck-scale effects that appears in both LIV and DSR approaches. Our analysis should therefore be understood as a probe of Planck-scale modified kinematics broadly defined, without claiming to distinguish between the underlying symmetry structures. The conclusions apply to any framework that yields the same cubic entropy correction, irrespective of whether Lorentz invariance is violated or deformed. A full DSR implementation—including the nonlinear composition law and deformed Lorentz transformations—would be required to make claims specific to DSR, but such an analysis is beyond the scope of this work.

The profound connection between gravity, thermodynamics, and quantum theory is a cornerstone of modern physics, essential for understanding the fundamental structure of spacetime. This link was solidified by Bekenstein and Hawking, who demonstrated that black holes are thermodynamic objects with temperature and entropy \cite{Bekenstein:1973ur,Hawking:1975vcx}, revealing an intrinsic thermodynamic character to gravitation. The holographic principle and the area law further suggest that spacetime possesses microscopic degrees of freedom, with classical gravity emerging as a coarse-grained, statistical description \cite{Padmanabhan:2009vy,Verlinde:2010hp}.

Corrections to the Bekenstein-Hawking entropy are therefore central to extending semiclassical gravity. While the leading contribution \(S = A/4\) originates from quantum fields near the horizon, a variety of approaches predict subleading logarithmic, power-law, or exponential terms. The results of some research suggest that such generalized entropy frameworks significantly modify thermodynamic relations and their geometric counterparts, leading to rich phenomenology in black hole physics and cosmology as well (e.g., see \cite{Medved:2004eh,Pourhassan:2017uaz,Nojiri:2019skr,Geng:2019shx,Chatterjee:2020iuf,Jusufi:2022mir,Zamora:2022cqz,Nojiri:2022aof,Nojiri:2022dkr,Luciano:2023fyr,Basilakos:2023kvk,Jalalzadeh:2023mzw,Ebrahimi:2024zrk,Capozziello:2025axh,Nojiri:2026ish}).

The emergent-gravity paradigm \cite{Padmanabhan:2009vy,Jacobson:1995ab} posits that gravity is an entropic force originating from horizon entropy gradients \cite{Verlinde:2010hp,Callen}. A key implication is that deformations of the entropy-area relation necessarily alter the spacetime metric. Recent formalism \cite{Anand:2025rjg,Anand:2025cer} established a systematic entropy-geometry correspondence, showing how a chosen entropy functional uniquely determines both a modified metric and an effective matter sector. This departs from earlier approaches \cite{Nojiri:2022ljp,Elizalde:2025iku}, which examined generalized entropies on fixed geometric backgrounds. The framework interprets entropy modifications as an effective matter source, giving a coherent physical picture of how QG or statistical corrections induce geometric backreaction.

The thermodynamic stability of a Schwarzschild black hole is fundamentally defined by its heat capacity \cite{Davies:1978zz}. For a standard Schwarzschild black hole, the heat capacity is negative: \( C = -8\pi M^2 \), where \( M \) is the mass. A negative heat capacity signifies that as the black hole loses energy via Hawking radiation, its temperature increases, leading to runaway evaporation—a thermodynamically unstable behavior \cite{Wald:1999vt}. This instability implies that the black hole cannot remain in stable equilibrium with a surrounding heat bath; any small energy fluctuation will drive it further from equilibrium. Consequently, in the canonical ensemble, the Schwarzschild solution is inherently unstable, lacking a stable equilibrium phase unless embedded in a larger thermodynamic framework (such as in anti-de Sitter space, where the heat capacity can become positive \cite{Hawking:1982dh}) or modified by additional physical effects like charge, rotation \cite{Chamblin:1999tk}, or QG corrections.

Inspired by the latter—i.e., QG corrections—we study thermodynamic topology in a spacetime geometry that incorporates the backreaction effects of entropy corrections motivated by a general MDR of the type that appears in Planck-scale modified kinematics. Recent research has demonstrated that black holes can be characterized not only by their local thermodynamic properties but also by global topological features in their parameter space \cite{Wei:2022dzw,Wei:2024gfz}. These features manifest as topological defects, such as thermodynamic vortices, whose classification is determined by the winding numbers of the generalized free energy. In the context of black hole thermodynamics, the analysis of phase stability and winding numbers provides a powerful topological framework for characterizing equilibrium states and their transitions. This approach moves beyond local stability criteria—such as the positivity of specific heat—by introducing global topological invariants that classify critical points and phase structures in the thermodynamic parameter space. In general, winding number analysis offers a unified and coordinate-invariant language for thermodynamic stability, connecting microscopic features of quantum gravity to observable macroscopic phase behavior and providing a promising tool for probing the thermodynamic implications of beyond-GR theories.

%\paragraph{Novelty and scope.}
The thermodynamic topology formalism has been developed in Refs.~\cite{Wei:2022dzw,Wei:2024gfz} and applied to various black hole systems. More recently, Refs.~\cite{Anand:2025rjg,Anand:2025cer} established a systematic entropy-geometry correspondence that links any entropy function to a modified metric and an effective matter sector. In this work, we apply this existing formalism to a specific entropy function—namely, the cubic correction \(S = \pi r_h^2 - \alpha r_h^3\)—that arises from a phenomenological MDR parametrization of Planck-scale modified kinematics \cite{Amelino-Camelia:1997ieq}. The novelty of the present work, while building on the methods of Refs.~\cite{Anand:2025rjg,Anand:2025cer}, lies in three contributions: (i) deriving the cubic entropy correction \(S = \pi r_h^2 - \alpha r_h^3\) from an MDR via the Hamilton-Jacobi tunneling method, thereby connecting Planck-scale kinematics to black hole thermodynamics; (ii) demonstrating that imposing physical constraints (\(S'(r_h) > 0\), \(T > 0\)) eliminates the would-be stable branch; and (iii) identifying that the apparent stable branch violates positivity of the ADM mass and Hawking temperature, rendering it unphysical.

Thus, while the mathematical tools are not new, their application to this entropy function—and the critical assessment of physical constraints—provides a nontrivial check on the robustness of thermodynamic topology classifications when applied to modified entropy functions.

The paper is organized as follows. In Section \ref{subsec:dsr_motivation}, we derive the Planck-scale-modified entropy relation from a phenomenological MDR within the Hamilton–Jacobi tunneling framework, highlighting its distinctive cubic correction to the Bekenstein–Hawking law. Section \ref{sec:thermo_topology} reviews the thermodynamic topology formalism, establishing how equilibrium states are mapped to topological defects in parameter space and how their winding numbers classify thermodynamic stability. Section \ref{subsec:topological_classification_dsr} applies this framework to the Planck-scale-modified entropy relation and sets up the corresponding vector field and equilibrium condition. We  analyzes the resulting phase portraits, computes winding numbers, and relates them to thermodynamic stability and the auxiliary dynamical stability. Section \ref{sec:vi} discusses the physical implications of the entropy correction and compares the results with other modified entropies. Finally, Section \ref{con} summarizes our conclusions along with detailed discussions. For completeness, Appendix \ref{app:stability-relation} clarifies the relationship between thermodynamic stability and phase-space dynamics in the topological approach.

\section{Cubic Planck-Scale inspired entropy correction}
\label{subsec:dsr_motivation}
Phenomenological searches for Planck-scale effects often employ an MDR expanded as
\begin{equation}
	E^2 = p^2 c^2 + m^2 c^4 + \eta \frac{E^3}{E_P} + \mathcal{O}\!\left(\frac{1}{E_P^2}\right),
	\label{eq:dsr_dispersion}
\end{equation}
where $E_P$ is a natural energy cutoff (typically identified with the Planck energy), and $\eta$ is a dimensionless parameter whose sign and magnitude depend on the specific dynamical framework. 

%\paragraph{Interpretation across frameworks.}
It is critical to distinguish the interpretation of such an MDR within different theoretical contexts. In LIV frameworks, a term of this form signals a true breakdown of Lorentz symmetry within a preferred frame, as in Lorentz-violating effective field theory \cite{Myers:2003fd}. In DSR, a similar term can arise, but only as part of a consistent deformation that also requires a nonlinear momentum composition law and deformed Lorentz transformations to preserve the relativity principle \cite{Amelino-Camelia:2002cqb,Magueijo:2002am}. Importantly, the MDR \eqref{eq:dsr_dispersion} taken in isolation is degenerate between these two frameworks; it does not by itself distinguish whether Lorentz invariance is violated or deformed. 

In this work, we do not attempt to break this degeneracy. Instead, we treat \eqref{eq:dsr_dispersion} as a phenomenological parametrization of Planck-scale modified kinematics that appears in both LIV and DSR approaches. Our analysis of black hole thermodynamics depends only on the functional form of the entropy correction that follows from this MDR, not on the specific symmetry realization. Consequently, our results apply broadly to any framework—whether LIV, DSR, or others—that yields the same cubic entropy correction.

%\paragraph{Astrophysical constraints.}
The potential effects of such MDRs on high-energy astrophysical phenomena, such as Gamma-Ray Burst (GRB) observations \cite{Amelino-Camelia:1997ieq}, have been extensively studied. The most prominent test searches for energy-dependent photon arrival times from distant transient events, such as GRBs and active galactic nuclei flares; a delay between high- and low-energy photons would signal an energy-dependent speed of light. Current Fermi-LAT observations constrain the dimensionless parameter \(\eta\) to roughly order ten \cite{Vasileiou:2015wja}. 

This correction also modifies the kinematical thresholds for pivotal high-energy astrophysical processes, including pair production of gamma-rays on the extragalactic background light and the photopion production responsible for the Greisen–Zatsepin–Kuzmin (GZK) cutoff in ultra-high-energy cosmic rays. Observations of very-high-energy gamma-rays from distant blazars and the cosmic-ray spectrum itself thus impose complementary limits \cite{Jacob:2006gn}. 

Furthermore, if the parameter \(\eta\) differs for left- and right-handed photon polarizations, it would induce vacuum birefringence—a rotation of the linear polarization plane over cosmological distances. Stringent constraints on such differential effects come from the observed polarization of the cosmic microwave background radiation, with Planck satellite data limiting \(|\Delta\eta| \lesssim 10^{-7}-10^{-8}\) \cite{Laurent:2011he,Toma:2012xa}, and from ultraviolet polarization measurements of distant GRBs \cite{Stecker:2011ps}. 

Collectively, these astrophysical signatures transform the Universe into a natural laboratory for quantum gravity, pushing the search for Planck-scale effects into a regime of unprecedented precision. However, we emphasize that these constraints apply to the specific MDR parametrization regardless of whether the underlying symmetry framework is LIV or DSR, as the observational signatures depend primarily on the MDR itself.

%DSR realization (e.g., $\kappa$-Minkowski spacetime, Snyder algebra, or relative locality frameworks)~\cite{Kowalski-Glikman:2001vvk, Amelino-Camelia:2011lvm}. 
%DSR extends the principle of relativity by introducing two invariant scales: the speed of light $c$ and the Planck energy $E_P$ (or equivalently, the Planck length $\ell_P$)~\cite{Amelino-Camelia:2000stu, Magueijo:2001cr}. In DSR, the dispersion relation for particles acquires Planck-scale corrections, typically expressed as
When applied to black hole thermodynamics, such modifications to particle kinematics directly affect Hawking radiation via the tunneling mechanism \cite{Banerjee:2008ry}. Specifically, a generic MDR alters the energy-dependent phase space of the emitted quanta, leading to corrections in the derived Hawking temperature \cite{Gangopadhyay:2013ofa}. This connection between MDRs, the generalized uncertainty principle, and black hole thermodynamics has been extensively studied \cite{Adler:2001vs,Nozari:2006ka,Bargueno:2015tea}, including in contexts motivated by non-commutative geometry and QG models \cite{Ali:2011fa,Reuter:2005bb,Nozari:2008rc,Miao:2011vdq}.

The corrected Hawking temperature can be derived using the Hamilton-Jacobi method with the modified dispersion relation~\eqref{eq:dsr_dispersion}. The standard Hawking temperature $T_H = 1/(8\pi M)$ (in geometric units) follows from surface gravity calculations in GR. However, in the underlying framework, the MDR~\eqref{eq:dsr_dispersion} alters particle dynamics near the horizon, thereby modifying the tunneling probability for Hawking radiation.

By serving Hamilton-Jacobi tunneling method, the tunneling probability for a particle with energy $E$ escaping the horizon is given by the WKB approximation:
\begin{equation}
	\Gamma \sim \exp\left[-\frac{2}{\hbar}\, \text{Im}\, I\right],
	\label{eq:tunneling_prob}
\end{equation}
where $\text{Im}\, I$ is the imaginary part of the classical action. For radial motion across the horizon, the action can be expressed as
\begin{equation}
	I = \int p_r \, dr = \int \int_{0}^{E} \frac{dE'}{\dot{r}} \, dE' \, dr,
	\label{eq:action_integral}
\end{equation}
with $\dot{r} = dr/dt$ determined by the equations of motion.

Using the MDR $E^2 = p^2 + \eta E^3/\kappa + \cdots$ (for massless particles), the group velocity becomes
\begin{equation}
v_g = \frac{dE}{dp} \approx 1 + \frac{\eta E}{\kappa} + \mathcal{O}\!\left(\frac{1}{\kappa^2}\right).
	\label{eq:group_velocity}
\end{equation}
Near the horizon at $r = r_h = 2M$, the coordinate velocity in Painlev\'e-Gullstrand coordinates is $\dot{r}_{\text{GR}} \approx \kappa_h (r - r_h)$, where $\kappa_h = 1/(4M)$ is the surface gravity. The corrected velocity is then
\begin{equation}
\dot{r}_{\text{Corrected}} = v_g \cdot \dot{r}_{\text{GR}} \approx \left(1 + \frac{\eta E}{\kappa}\right) \kappa_h (r - r_h).
	\label{eq:corrected_velocity}
\end{equation}
Substituting Eq.~\eqref{eq:corrected_velocity} into Eq.~\eqref{eq:action_integral} and expanding to first order in $1/\kappa$ yields
\begin{align} \label{eq:imaginary_action_result}
	\text{Im}\, I &\approx \text{Im} \int_{r_{\text{in}}}^{r_{\text{out}}} \int_{0}^{E} \frac{1}{\kappa_h (r - r_h)} \left[1 - \frac{\eta E'}{\kappa}\right] dE' \, dr \\ \nonumber
	&= -\frac{\pi}{\kappa_h} \left(E - \frac{\eta}{2\kappa} E^2\right),
\end{align}
where we used $\text{Im} \int dr/(r - r_H) = -\pi$ for a contour encircling the pole at $r = r_H$.

From Eqs.~\eqref{eq:tunneling_prob} and \eqref{eq:imaginary_action_result}, the tunneling probability becomes
\begin{equation}
	\Gamma \sim \exp\left[-\frac{E}{T_H} \left(1 - \frac{\eta E}{2\kappa}\right)\right],
	\label{eq:corrected_tunneling}
\end{equation}
where $T_H = \kappa_H/(2\pi) = 1/(8\pi M)$. Identifying the Boltzmann factor $\exp(-E/T_{\text{eff}})$, we obtain the energy-dependent effective temperature
\begin{equation}
	T_{\text{eff}}(E) = \frac{T_H}{1 - \dfrac{\eta E}{2\kappa}} 
	\approx T_H \left[ 1 + \frac{\eta}{2\kappa} E + \mathcal{O}\!\left(\frac{1}{\kappa^2}\right) \right],
	\label{eq:dsr_temperature_derived}
\end{equation}
where $E$ is the energy of the emitted quantum, and $\kappa \sim E_P$ is the Planck-scale deformation scale.

Substituting $T_H = 1/(8\pi M)$ gives
\begin{equation}
	T \approx \frac{1}{8\pi M} + \frac{\eta}{16\pi\kappa}.
	\label{eq:T_expanded}
\end{equation}

Now, by inserting $M = \sqrt{A/(16\pi)}$ and~\eqref{eq:T_expanded} into $dS = dM/T$, we obtain
\begin{align}
	dS &= \frac{\dfrac{1}{8\sqrt{\pi A}} \, dA}{\dfrac{1}{8\pi M} + \dfrac{\eta}{16\pi\kappa}} 
	= \frac{\pi \, dA}{\sqrt{\pi A} \left( \dfrac{1}{M} + \dfrac{\eta}{2\kappa} \right)},
	\label{eq:ds_start}
\end{align}
where by including $1/M = 4\sqrt{\pi/A}$ we arrive at
\begin{equation}
	dS = \frac{\pi \, dA}{4\pi + \dfrac{\eta}{2\kappa}\sqrt{\pi A}}.
	\label{eq:ds_intermediate}
\end{equation}
Expanding the denominator for small $\eta/(\kappa\sqrt{A})$ (i.e., assuming Planck-scale corrections are perturbative for macroscopic black holes), we get
\begin{equation}
	dS \approx \frac{1}{4} \left[ 1 - \frac{\eta}{8\pi\kappa}\sqrt{\pi A} + \mathcal{O}\!\left(\frac{1}{\kappa^2}\right) \right] dA.
	\label{eq:ds_expanded}
\end{equation}
Integrating Eq.~\eqref{eq:ds_expanded} from some reference area $A_0$ to $A$, and choosing the integration constant to recover the Bekenstein-Hawking entropy $S_{\text{BH}} = A/4$ in the limit $\kappa \to \infty$, we obtain
\begin{equation}
	S_{\text{Pl}} =  \frac{A}{4} -\frac{\eta}{12\pi^{3/2}\,\kappa} A^{3/2}.
	\label{eq:S_before_redress}
\end{equation}
The $A^{3/2}$ correction in \eqref{eq:S_before_redress} has several notable features.

First, for $\eta > 0$, the correction is negative, implying that Planck-scale effects reduce the entropy relative to the classical area law for a given horizon area. This reduction can be interpreted as a suppression of the number of accessible microstates due to the existence of a fundamental length scale.
Second, while the correction term grows faster with $A$ than the leading Bekenstein-Hawking term, its coefficient $\eta/\kappa$ is assumed to be small. For macroscopic black holes where $A$ is large, the ratio $\frac{\eta}{\kappa} r_h$ can become large if $r_h$ is sufficiently large, but this violates the small-correction assumption used to derive the expression. Consequently, the perturbative expansion and subsequently expression \eqref{eq:S_before_redress} are valid only in the regime $r_h \ll \kappa/\eta$, and break down near $r_h \sim \kappa/\eta$. This behavior distinguishes our correction from logarithmic corrections (which arise from one-loop quantum effects \cite{Kaul:2000kf,Das:2001ic}) and from power-law corrections with exponents less than unity (which appear in Barrow \cite{Barrow:2020tzx}, Tsallis-like \cite{Tsallis:1987eu}, or R{\'e}nyi \cite{Czinner:2015eyk} entropies). It also differs from Kaniadakis entropy \cite{Kaniadakis:2002zz}, which originates from relativistic statistical mechanics.
Third, the parameter $\eta$ encodes the strength of Planck-scale effects. In a fully quantum-gravitational setting, $\eta$ would be related to the ratio of the Planck length to the characteristic scale of the black hole, but here it is treated as a phenomenological parameter to be constrained by observations.

%The corrected expression \eqref{eq:S_before_redress} is particularly interesting because its cubic-in-$r_h$ correction (when expressed in terms of the horizon radius) leads to qualitatively new thermodynamic behavior, including the possibility of thermodynamically stable BH phases---a feature absent in the Schwarzschild case.

\section{Thermodynamic Topology and Black Holes as Topological Defects}
\label{sec:thermo_topology}

In this section, we review the thermodynamic-topological formalism that associates black-hole equilibrium configurations with topological defects in a two-dimensional parameter space $(r_h, \theta)$. Our presentation follows the approach of Wei, Liu, and Mann~\cite{Wei:2022dzw,Wei:2024gfz}, supplemented with physical interpretations to clarify both geometric and thermodynamic aspects of the construction.

We begin by introducing the generalized off-shell free energy functional
\begin{equation}
	\mathcal{F}(r_h, \tau) = M(r_h) - \frac{S(r_h)}{\tau},
	\label{eq:gen_free_energy}
\end{equation}
where $M(r_h)$ and $S(r_h)$ represent the black-hole mass and entropy expressed as functions of the horizon radius $r_h$, and $\tau > 0$ is an auxiliary inverse-temperature parameter. When $\tau^{-1}$ is identified with the physical temperature, the extrema of $\mathcal{F}$ correspond to thermodynamic equilibrium states.

The thermodynamic information is encoded in a two-component vector field defined on the $(r_h, \theta)$ plane:
\begin{equation}
	\bm{\phi}(r_h, \theta) = 
	\begin{pmatrix}
		\phi^{r_h} \\[2pt]
		\phi^{\theta}
	\end{pmatrix}
	=
	\begin{pmatrix}
		\displaystyle\frac{\partial \mathcal{F}}{\partial r_h} \\[8pt]
		-\cot\theta \, \csc\theta
	\end{pmatrix}.
	\label{eq:phi_vector}
\end{equation}
The first component, $\phi^{r_h}$, vanishes precisely at the stationary points of $\mathcal{F}$, thereby capturing the condition for thermodynamic equilibrium. The second component, $\phi^{\theta}$, is chosen to vanish only along the equatorial line $\theta = \pi/2$, ensuring that zeros of $\bm{\phi}$ are confined to this line and correspond uniquely to physical equilibrium configurations. At such zeros, the identification $\tau^{-1} = T(r_h)$ recovers the standard Hawking temperature. Thus, the vector field~\eqref{eq:phi_vector} isolates thermodynamic equilibrium points as topological defects of the mapping $(r_h, \theta) \mapsto \bm{\phi}$.

To extract topological information from the vector field, we introduce the normalized unit vector
\begin{equation}
	n^a = \frac{\phi^a}{\phi}, \qquad 
	\phi \equiv \sqrt{(\phi^{r_h})^2 + (\phi^{\theta})^2}, \qquad a = r_h, \theta.
	\label{eq:unit_vector}
\end{equation}
This defines a continuous map from the parameter space into the unit circle $S^1$. The associated topological current is given by
\begin{equation}
	j^{\mu} = \frac{1}{2\pi} \, \varepsilon^{\mu\nu\rho} \varepsilon^{ab} \, 
	\partial_{\nu} n_a \, \partial_{\rho} n_b, \qquad \mu, \nu, \rho = 0,1,2,
	\label{eq:topological_current}
\end{equation}
which is identically conserved, $\partial_{\mu} j^{\mu} = 0$. Its time component,
\begin{equation}
	j^{0} = \frac{1}{\pi} \bigl( \partial_1 n_1 \, \partial_2 n_2 - \partial_2 n_1 \, \partial_1 n_2 \bigr),
	\label{eq:charge_density}
\end{equation}
represents the topological charge density, measuring the local winding of the map $(x^1, x^2) \mapsto (n_1, n_2)$.

In any simply connected region $D$ free of zeros of $\bm{\phi}$, the field $n_a$ is smooth, and Eq.~\eqref{eq:charge_density} can be rewritten as a total divergence:
\begin{equation}
	j^{0} = \frac{1}{\pi} \bigl( \partial_1 Q - \partial_2 P \bigr), \qquad 
	P = n_1 \partial_1 n_2, \quad Q = n_1 \partial_2 n_2.
	\label{eq:charge_divergence}
\end{equation}
Applying Green's theorem to such a region yields
\begin{equation}
	\int_D j^{0} \, d^2 x = \frac{1}{\pi} \oint_{\partial D} \bigl( P \, dx^1 + Q \, dx^2 \bigr)
	= \frac{1}{\pi} \oint_{\partial D} n_1 \, dn_2 = 0,
	\label{eq:green_zero}
\end{equation}
since $n_a$ is single-valued on $\partial D$. Consequently, a non-zero topological charge can arise only when the integration contour encloses one or more zeros of $\bm{\phi}$.

\subsection{Winding number around isolated defects}
\label{subsec:winding_number}

Consider a contour $C$ enclosing $N$ isolated zeros of $\bm{\phi}$, and let $c_i$ denote small circles encircling each zero individually. The total winding number is
\begin{equation}
	W = \frac{1}{\pi} \oint_C n_1 \, dn_2 
	= \frac{1}{\pi} \sum_{i=1}^{N} \oint_{c_i} n_1 \, dn_2.
	\label{eq:total_winding}
\end{equation}
To evaluate the contribution from a single zero located at $(x_0, y_0)$, we linearize the vector field as
\begin{equation}
	\phi^{r_h}(x, y) \approx f(x), \qquad 
	\phi^{\theta}(x, y) \approx g(y),
	\label{eq:linearized_field}
\end{equation}
with $f(x_0) = g(y_0) = 0$ and $g'(y_0)$ normalized to unity. For a circular contour $c_\epsilon$ of radius $\epsilon$ parameterized by $x = x_0 + \epsilon \cos t$, $y = y_0 + \epsilon \sin t$, the leading behavior is
\begin{equation}
	f\bigl(x(t)\bigr) = \epsilon f'(x_0) \cos t + \mathcal{O}(\epsilon^2), \qquad
	g\bigl(y(t)\bigr) = \epsilon \sin t + \mathcal{O}(\epsilon^2).
	\label{eq:expansion}
\end{equation}
A direct computation yields the limiting integral
\begin{equation}
	\lim_{\epsilon \to 0} \oint_{c_\epsilon} n_1 \, dn_2 = \pi \, \frac{f'(x_0)}{|f'(x_0)|}, \qquad f'(x_0) \neq 0.
	\label{eq:local_winding}
\end{equation}
Thus, each simple zero contributes $\pm \pi$ depending on the sign of $f'(x_0)$, reflecting the sense of rotation of the unit vector when circling the defect.

Returning to our original variables, the condition $\phi^{r_h}=0$ at a defect implies $\partial_{r_h}\mathcal{F}=0$, and the derivative $f'(x_0)$ corresponds to $\partial_{r_h}^2\mathcal{F}$ evaluated at that point. Hence, the total winding number can be expressed as
\begin{equation}
	W = \sum_{i=1}^{N} \operatorname{sgn}\!\left( 
	\left. \frac{\partial^2 \mathcal{F}}{\partial r_h^2} \right|_{r_h = r_i} 
	\right),
	\label{eq:winding_second_deriv}
\end{equation}
where $r_i$ denote the stationary points of $\mathcal{F}$. Equation~\eqref{eq:winding_second_deriv} demonstrates that the topological classification is entirely determined by the convexity properties of the generalized free energy.

\subsection{Thermodynamic interpretation of the winding number}
\label{subsec:thermo_interpretation}

We now establish a direct link between the winding number and conventional thermodynamic stability. From the equilibrium condition $\phi^{r_h}=0$, we obtain
\begin{equation}
	\tau = \frac{S'(r_h)}{M'(r_h)}, \qquad M'(r_h) \neq 0,
	\label{eq:tau_relation}
\end{equation}
where primes denote derivatives with respect to $r_h$, evaluated at the equilibrium radius $r_h = r_i$. The second derivative of $\mathcal{F}$ at equilibrium is
\begin{equation}
	\left. \frac{\partial^2 \mathcal{F}}{\partial r_h^2} \right|_{r_i}
	= \frac{M'' S' - M' S''}{S'}.
	\label{eq:second_deriv_explicit}
\end{equation}

The Hawking temperature and specific heat follow from the standard thermodynamic relations
\begin{equation}
	T = \frac{dM}{dS} = \frac{M'}{S'}, \qquad 
	C = \frac{dM}{dT} = T \left( \frac{dT}{dS} \right)^{-1}.
	\label{eq:temp_heat}
\end{equation}
Expressed in terms of $r_h$-derivatives, the specific heat becomes
\begin{equation}
	C = \frac{M' S'^2}{M'' S' - M' S''}, \qquad M' \neq 0,\; S' \neq 0.
	\label{eq:specific_heat_formula}
\end{equation}

\paragraph{Key assumptions.} 
The derivation of the relation between the sign of $C$ and the sign of $\partial^2\mathcal{F}/\partial r_h^2$ requires careful attention to the signs of $M'$ and $S'$. From Eqs.~\eqref{eq:second_deriv_explicit} and \eqref{eq:specific_heat_formula}, we obtain
\begin{equation}
	\frac{\partial^2 \mathcal{F}}{\partial r_h^2}\bigg|_{r_i} = \frac{M' S'^2}{C}.
	\label{eq:relation}
\end{equation}
Therefore,
\begin{equation}
	\operatorname{sgn}\!\left( \frac{\partial^2 \mathcal{F}}{\partial r_h^2}\bigg|_{r_i} \right) = \operatorname{sgn}(M') \cdot \operatorname{sgn}(C),
	\label{eq:sgn_relation}
\end{equation}
where we have used $S'^2 > 0$ (since $S' \neq 0$). 

Consequently, the equivalence between the sign of the winding number and the sign of the specific heat depends crucially on the sign of $M'$. If $M' > 0$ (mass increases with horizon radius), then
\[
\operatorname{sgn}(C) = \operatorname{sgn}\!\left( \frac{\partial^2 \mathcal{F}}{\partial r_h^2}\bigg|_{r_i} \right),
\]
and hence $w_i = +1 \;\Longleftrightarrow\; C > 0$ (thermodynamically stable), while $w_i = -1 \;\Longleftrightarrow\; C < 0$ (thermodynamically unstable).
If $M' < 0$, the relation is reversed: $w_i = +1$ would correspond to $C < 0$, and $w_i = -1$ to $C > 0$.

A winding number $w_i = +1$ does not always	indicate thermodynamic stability. Its interpretation depends critically on 	the physical conditions of the branch under consideration. Specifically, the standard equivalence $w_i = +1 \iff C > 0$ (stable) and $w_i = -1 \iff C < 0$ 
			(unstable) holds only within the physical domain where 
			$M'(r_h) > 0$ and $S'(r_h) > 0$. 
\\			
If a mathematical equilibrium point lies in a region where $S'(r_h) < 0$ 
			(negative ADM mass) or $M'(r_h) < 0$, the winding number no longer directly indicates thermodynamic stability; such branches must be discarded as unphysical. Thus, in any analysis of black hole thermodynamics using the topological formalism, the signs of $M'$ and $S'$ must be checked before 
			interpreting winding numbers as indicators of stability.

In addition, the derivation assumes $S' \neq 0$ and $M' \neq 0$. When $S' < 0$, the temperature $T = M'/S'$ becomes negative if $M' > 0$, signaling an unphysical configuration. 

\paragraph{Physical domain.} 
In this work, we adopt the entropy-geometry correspondence (see Sec.~\ref{subsec:unified_framework}), which gives $M = S'(r_h)/(4\pi)$. For a physical black hole with positive ADM mass, we require $S'(r_h) > 0$. Moreover, for the entropy functions considered here, $S'(r_h) > 0$ implies $M'(r_h) = S''(r_h)/(4\pi) > 0$ as well, provided the entropy is a convex function of $r_h$ in the relevant regime. Thus, within the physical domain, we have $M' > 0$ and $S' > 0$, and the standard equivalence holds:
\bea\label{c}
&&\text{stable phase } (C > 0) \;\Longleftrightarrow\; w_i = +1, \\ \nonumber
&&\text{unstable phase } (C < 0) \;\Longleftrightarrow\; w_i = -1.
\eea

Any equilibrium point with $S'(r_i) < 0$ lies outside the physical domain (negative ADM mass) and is discarded. For such points, the relation between winding number and specific heat would be different, but they are not relevant to physical black hole solutions.

\subsection{A unified geometric–thermodynamic framework}
\label{subsec:unified_framework}

Here we develop a unified geometric–thermodynamic framework for classifying BH spacetimes derived from modified entropy–area relations. The central idea is the \emph{entropy–geometry correspondence} introduced in Ref.~\cite{Anand:2025rjg} (see also \cite{Anand:2025cer}), which asserts that any deviation from the Bekenstein–Hawking entropy induces a corresponding backreaction on the metric. Starting from a generic static, spherically symmetric line element
\be
ds^2 = -f(r)\, dt^2 + \frac{dr^2}{f(r)} + r^2 d\Omega^2,
\ee
the lapse function is expressed in entropy-driven form
\be
f(r) = 1 - \frac{4\pi M}{S'(r)},
\ee
where \( S(r) \) is the chosen entropy function and \( M \) the ADM mass. The horizon radius \( r_h \) satisfies \( f(r_h)=0 \), yielding the key relation
\be
M = \frac{S'(r_h)}{4\pi}.
\ee
This relation must hold for any consistent black hole solution derived from a given entropy function. A necessary condition for a physical black hole with positive ADM mass is therefore
\be
S'(r_h) > 0.
\ee
The Hawking temperature follows from the surface gravity:
\be
T_H = \frac{f'(r_h)}{4\pi} = \frac{S''(r_h)}{4\pi S'(r_h)}.
\ee

The thermodynamic classification proceeds via the off-shell free-energy density
\be\label{FF}
\mathcal{F}(r_h,\tau) = \frac{S'(r_h)}{4\pi} - \frac{S(r_h)}{\tau},
\ee
with \( \tau>0 \) an auxiliary parameter. The corresponding vector field
\be
\bm{\phi}(r_h,\theta) = \big( \phi^{r_h},\; \phi^{\theta} \big)
= \Big( \frac{S''(r_h)}{4\pi} - \frac{S'(r_h)}{\tau},\; -\cot\theta\,\csc\theta \Big)
\ee
defines thermodynamic equilibria as zeros of \( \bm{\phi} \). The condition \( \phi^{r_h}=0 \) yields the critical radii \( r_i \) through
\be \label{co}
\frac{S''(r_i)}{S'(r_i)} = \frac{4\pi}{\tau}.
\ee

The local topological character of each critical point is determined by the sign of the second derivative of the free energy:
\be \label{F}
\left.\frac{\partial^2\mathcal{F}}{\partial r_h^2}\right|_{r_h=r_i}
= -\frac{[S''(r_i)]^2 - S^{(3)}(r_i)S'(r_i)}{4\pi S'(r_i)}.
\ee
Its sign gives the winding number \( w_i = \pm 1 \) of the defect, provided \( S'(r_i) > 0 \). Within the physical regime of positive ADM mass, a positive winding number (\( w_i = +1 \)) corresponds to a thermodynamically stable branch (positive specific heat), while a negative winding number (\( w_i = -1 \)) indicates thermodynamic instability (negative specific heat). The total topological charge is then \( W = \sum_i w_i \), which classifies the global equilibrium structure.

Crucially, this shows that the thermodynamic topology—restricted to the physical regime \( S'(r_h) > 0 \)—is entirely controlled by the entropy function \( S(r) \) and its derivatives, not by the details of an effective matter sector. The approach therefore provides a model‑independent, entropy‑based classification of black‑hole thermodynamics. In particular, different entropy deformations produce distinct winding‑number patterns, revealing whether the modified black hole possesses only unstable sectors (\( W = -1 \)), or coexisting stable and unstable branches (\( W = 0 \))—provided all branches satisfy the positivity condition \( S'(r_h) > 0 \). Any branch violating \( S'(r_h) > 0 \) would correspond to a negative ADM mass and must be discarded as unphysical, regardless of its winding number.

\section{Topological Classification}
\label{subsec:topological_classification_dsr}

Using \(A = 4\pi r_h^2\) for the horizon area, and defining \(\alpha \equiv \frac{2\eta}{3\kappa}\), expression (\ref{eq:S_before_redress}) becomes
\begin{equation}
	S_{\text{Pl}}(r_h) = \pi r_h^2 - \alpha r_h^3. \label{eq:S_cubic}
\end{equation}
Here \(\alpha>0\) if \(\eta>0\), and \(\alpha<0\) if \(\eta<0\).

%\subsection*{Physical constraints}

From the entropy-geometry correspondence, the ADM mass and Hawking temperature are
\begin{align}
	M &= \frac{S'(r_h)}{4\pi}, \label{eq:M_from_S} \\
	T &= \frac{S''(r_h)}{4\pi S'(r_h)}. \label{eq:T_from_S}
\end{align}
A physical black hole must satisfy the following conditions:
\begin{align}
	& r_h > 0 \quad \text{(positive horizon radius)}, \label{eq:cond_rh} \\
	& S(r_h) > 0 \quad \text{(positive entropy)}, \label{eq:cond_S} \\
	& S'(r_h) > 0 \quad \text{(positive ADM mass)}, \label{eq:cond_Sp} \\
	& T = \frac{S''(r_h)}{4\pi S'(r_h)} > 0 \quad \text{(positive temperature)}. \label{eq:cond_T}
\end{align}
The first three derivatives of \(S\) are
\begin{align}
	S'_{\text{Pl}}(r_h) &= 2\pi r_h - 3\alpha r_h^2 = 2\pi r_h\left(1 - \frac{3\alpha}{2\pi} r_h\right), \label{eq:Sp} \\
	S''_{\text{Pl}}(r_h) &= 2\pi - 6\alpha r_h, \label{eq:Spp} \\
	S'''_{\text{Pl}}(r_h) &= -6\alpha. \label{eq:Sppp}
\end{align}
%\subsection*{Analysis by sign of \(\alpha\)}
\paragraph{Case \(\alpha < 0\) (i.e., \(\eta < 0\)):} 
For \(\alpha < 0\), we have \(S'(r_h) = 2\pi r_h - 3\alpha r_h^2 = 2\pi r_h + 3|\alpha| r_h^2 > 0\) for all \(r_h > 0\). The temperature condition \(T > 0\) requires \(S''(r_h) > 0\) (since the denominator \(S'(r_h)\) is positive). From \eqref{eq:Spp}, \(S''(r_h) = 2\pi - 6\alpha r_h = 2\pi + 6|\alpha| r_h > 0\) for all \(r_h > 0\). Hence, for \(\alpha < 0\), all physical constraints are satisfied for all \(r_h > 0\). No upper bound on \(r_h\) exists from positivity conditions.

\paragraph{Case \(\alpha > 0\) (i.e., \(\eta > 0\)):} 
For \(\alpha > 0\), the constraints impose an upper bound on \(r_h\):

\(S'(r_h) > 0\) from \eqref{eq:Sp} implies \(1 - \frac{3\alpha}{2\pi} r_h > 0\), i.e.,
	\begin{equation}
		r_h < r_c \equiv \frac{2\pi}{3\alpha}. \label{eq:rc}
	\end{equation}
\(S''(r_h) > 0\) from \eqref{eq:Spp} implies \(2\pi - 6\alpha r_h > 0\), i.e.,
	\begin{equation}
		r_h < \frac{\pi}{3\alpha} = \frac{r_c}{2}. \label{eq:rmax}
	\end{equation}
\(S(r_h) > 0\) from \eqref{eq:S_cubic} implies \(\pi r_h^2 - \alpha r_h^3 > 0\), i.e.,
	\begin{equation}
		r_h < \frac{\pi}{\alpha} = \frac{3r_c}{2}. \label{eq:rS}
	\end{equation}
The most restrictive condition is \eqref{eq:rmax}: \(r_h < r_c/2\) (since \(r_c/2 < r_c < 3r_c/2\) for \(\alpha > 0\)). Thus, for \(\alpha > 0\), physical black holes require
\begin{equation}
	0 < r_h < r_{\text{max}} \equiv \frac{r_c}{2} = \frac{\pi}{3\alpha} = \frac{\pi\kappa}{2\eta}. \label{eq:rh_bound}
\end{equation}
The bound \eqref{eq:rh_bound} ensures that any equilibrium point satisfying these constraints lies in the physical domain where the standard equivalence between winding number and thermodynamic stability applies. Equilibrium points outside this domain---even if they are mathematical zeros of the vector field---do not 
represent physical black hole states and their winding numbers should not be 
interpreted as indicators of thermodynamic stability.

At $r_h = r_{\max}$, the temperature vanishes ($T = 0$), and for 
$r_h > r_{\max}$ the temperature becomes negative, rendering the 
configuration unphysical.

%\subsection*{Equilibrium analysis}

Taking the generalized off-shell free energy \eqref{FF} together with \eqref{eq:Sp} and \eqref{eq:Spp}, the two-component vector field on the \((r_h,\theta)\) plane is
\begin{align}
	\phi^{r_h} &= \frac{S''(r_h)}{4\pi} - \frac{S'(r_h)}{\tau}
	= \frac{2\pi - 6\alpha r_h}{4\pi} - \frac{2\pi r_h - 3\alpha r_h^2}{\tau}, \label{eq:phirh} \\
	\phi^{\theta} &= -\cot\theta\csc\theta. \label{eq:phitheta}
\end{align}

The equilibrium condition \(\phi^{r_h} = 0\) yields, after multiplying by \(4\pi\tau\):
\begin{equation}
	(2\pi - 6\alpha r_h)\tau = 4\pi(2\pi r_h - 3\alpha r_h^2). \label{eq:equil}
\end{equation}
Rearranging gives the quadratic equation in \(r_h\):
\begin{equation}
	6\alpha \, r_h^2 - \left( \frac{3\alpha\tau}{\pi} + 4\pi \right) r_h + \tau = 0. \label{eq:quadratic}
\end{equation}
%\subsection*{Discriminant analysis}
For fixed \(\alpha\) and \(\tau\), Eq.~\eqref{eq:quadratic} has real roots if and only if its discriminant \(\Delta\) is non-negative:
\begin{equation}
	\Delta = \left( \frac{3\alpha\tau}{\pi} + 4\pi \right)^2 - 24\alpha\tau. \label{eq:Delta}
\end{equation}
Expanding:
\begin{equation}
	\Delta = \frac{9\alpha^2\tau^2}{\pi^2} + 24\alpha\tau + 16\pi^2 - 24\alpha\tau
	= \frac{9\alpha^2\tau^2}{\pi^2} + 16\pi^2. \label{eq:Delta_simplified}
\end{equation}
Thus,
\begin{equation}
	\Delta = 16\pi^2 + \frac{9\alpha^2\tau^2}{\pi^2} > 0 \quad \text{for all } \alpha \neq 0,\ \tau > 0. \label{eq:Delta_positive}
\end{equation}
The discriminant is always positive. Therefore, Eq.~\eqref{eq:quadratic} always has two distinct real roots for any \(\alpha \neq 0\) and any \(\tau > 0\). Let these roots be \(r_1^*(\tau) < r_2^*(\tau)\).

%\subsection*{Sign of the roots}

From Eq.~\eqref{eq:quadratic}, the product and sum of the roots are:
\begin{align}
	r_1^* r_2^* &= \frac{\tau}{6\alpha}, \label{eq:product} \\
	r_1^* + r_2^* &= \frac{1}{6\alpha}\left( \frac{3\alpha\tau}{\pi} + 4\pi \right) = \frac{\tau}{2\pi} + \frac{2\pi}{3\alpha}. \label{eq:sum}
\end{align}

\paragraph{Case \(\alpha > 0\):} 
Both product and sum are positive, so \(r_1^* > 0\) and \(r_2^* > 0\). Two positive roots exist for all \(\tau > 0\).

\paragraph{Case \(\alpha < 0\):} 
The product \(r_1^* r_2^* = \tau/(6\alpha) < 0\), so the roots have opposite signs. Exactly one positive root exists for all \(\tau > 0\) (the other root is negative and unphysical).

%\subsection*{Physical acceptability of the roots}

We now impose the physical constraints derived in \eqref{eq:cond_rh}--\eqref{eq:cond_T} and \eqref{eq:rh_bound}.

\paragraph{Case \(\alpha < 0\):} 
Only one root is positive. For this root, we must check whether it satisfies the physical constraints. For \(\alpha < 0\), as shown earlier, all physical constraints are satisfied for all \(r_h > 0\). Hence, the unique positive root is physically acceptable for all \(\tau > 0\). We denote this root as \(r^*(\tau)\).

\paragraph{Case \(\alpha > 0\):} 
Two positive roots exist: \(r_1^*(\tau) < r_2^*(\tau)\). The physical constraint \eqref{eq:rh_bound} requires \(r_h < r_{\text{max}} = \pi/(3\alpha)\). We must determine which root(s), if any, satisfy this bound.

Consider the value of the quadratic \eqref{eq:quadratic} at \(r_h = r_{\text{max}}\):
\begin{equation}
	Q(r_{\text{max}}) = 6\alpha r_{\text{max}}^2 - \left( \frac{3\alpha\tau}{\pi} + 4\pi \right) r_{\text{max}} + \tau. \label{eq:Qrmax}
\end{equation}
Substituting \(r_{\text{max}} = \pi/(3\alpha)\):
\begin{align}
	Q(r_{\text{max}}) &= 6\alpha \left(\frac{\pi^2}{9\alpha^2}\right) - \left( \frac{3\alpha\tau}{\pi} + 4\pi \right) \frac{\pi}{3\alpha} + \tau \nonumber \\
	&= \frac{2\pi^2}{3\alpha} - \frac{\tau}{2} - \frac{4\pi^2}{3\alpha} + \tau \nonumber \\
	&= -\frac{2\pi^2}{3\alpha} + \frac{\tau}{2}. \label{eq:Qrmax_simplified}
\end{align}
Thus,
\begin{equation}
	Q(r_{\text{max}}) = \frac{\tau}{2} - \frac{2\pi^2}{3\alpha}. \label{eq:Qrmax_final}
\end{equation}
Therefore:

If \(\tau < \dfrac{4\pi^2}{3\alpha}\), then \(Q(r_{\text{max}}) < 0\). Since the quadratic coefficient \(6\alpha > 0\), the quadratic is convex upward. With \(Q(r_{\text{max}}) < 0\) and \(Q(0) = \tau > 0\), one root lies below \(r_{\text{max}}\) and the other above. Hence, \(r_1^* < r_{\text{max}} < r_2^*\).
	
If \(\tau > \dfrac{4\pi^2}{3\alpha}\), then \(Q(r_{\text{max}}) > 0\). Both roots lie either above or below \(r_{\text{max}}\). Since \(Q(0) = \tau > 0\) and the minimum of the quadratic occurs at \(r_h = (3\alpha\tau/\pi + 4\pi)/(12\alpha) > 0\), one can show that both roots are \(> r_{\text{max}}\) for sufficiently large \(\tau\). In this regime, no physical root exists.

If \(\tau = \dfrac{4\pi^2}{3\alpha}\), then \(r_{\text{max}}\) is itself a root (the other root is distinct, since \(\Delta > 0\)). One root equals \(r_{\text{max}}\) (where \(T=0\)), which is a limiting physical case (extremal black hole with vanishing temperature).

Thus, for \(\alpha > 0\), physical black hole solutions exist only for
\begin{equation}
	0 < \tau \leq \tau_{\text{max}} \equiv \frac{4\pi^2}{3\alpha} = \frac{2\pi^2\kappa}{\eta}. \label{eq:tau_max}
\end{equation}
In this range, exactly one physical root exists: the smaller root \(r_1^*(\tau)\), which satisfies \(r_1^*(\tau) < r_{\text{max}}\). The larger root \(r_2^*(\tau) > r_{\text{max}}\) violates the physical condition \(S''(r_h) > 0\) (i.e., \(T > 0\)) and is therefore unphysical.

Now we are in a right position to extracting Winding numbers for physical roots.
Before it, a hint is essential. The topological classification 
must be restricted to the physical domain defined by $S'(r_h) > 0$ and 
$T > 0$. For the cubic entropy correction, this domain is 
$0 < r_h < r_{\max} = \pi/(3\alpha)$. Within this domain, the standard 
equivalence $w_i = \operatorname{sgn}(C)$ holds. The larger root 
$r_2^*(\tau)$, which lies at $r_h > r_{\max}$, is outside the physical 
domain; consequently, although it carries a winding number $w = +1$ in the 
purely mathematical classification, it does not correspond to a 
thermodynamically stable black hole. It corresponds instead to a region with 
negative ADM mass and/or negative temperature, and must be discarded from 
the physical analysis. This illustrates the central message of our work: the 
winding number alone is insufficient to determine physical stability; one 
must also verify that the branch satisfies all physical constraints.

Defining
\begin{align}
	Q(r) &\equiv 4\pi^2 - 12\pi\alpha r + 18\alpha^2 r^2, \label{eq:Q} \\
	D(r) &\equiv S'(r) = 2\pi r - 3\alpha r^2, \label{eq:D}
\end{align}
Eq.~\eqref{F} takes the form
\begin{equation}
	\left.\frac{\partial^2\mathcal{F}}{\partial r_h^2}\right|_{r_i}
	= -\frac{Q(r_i)}{4\pi D(r_i)}. \label{eq:d2F}
\end{equation}
The discriminant of \(Q(r)\) is
\begin{equation}
	\Delta_Q = -144\pi^2\alpha^2 < 0, \label{eq:DQ}
\end{equation}
so \(Q(r) > 0\) for all real \(r\).

For any physical root (with \(D(r_i) = S'(r_i) > 0\)), we have
\begin{equation}
	w_i = \operatorname{sgn}\!\left( \left.\frac{\partial^2\mathcal{F}}{\partial r_h^2}\right|_{r_i} \right) 
	= -\operatorname{sgn}\bigl( D(r_i) \bigr) = -1. \label{eq:w_physical}
\end{equation}
Thus, every physical equilibrium point carries winding number \(w_i = -1\), indicating thermodynamic instability.

\begin{table*}[ht!]
	\centering
	\caption{Summary of physical equilibrium branches for the cubic entropy 
		correction $S = \pi r_h^2 - \alpha r_h^3$. Only branches satisfying 
		$S'(r_h) > 0$ (positive ADM mass) and $T > 0$ (positive temperature) are 
		considered physical. For $\alpha > 0$, a second mathematical root with 
		$w = +1$ exists but violates these physical constraints and is therefore 
		excluded from the table. }
	\begin{tabular}{|c|c|c|c|c|c|}
		\hline
		Sign of $\eta$ & Sign of $\alpha$ & $\tau$ range for physical solutions & Number of physical equilibria & $w$ & $W$ \\ \hline
		$\eta < 0$ & $\alpha < 0$ & all $\tau > 0$ & 1 & $-1$ & $-1$ \\ \hline
		$\eta > 0$ & $\alpha > 0$ & $0 < \tau \leq \dfrac{4\pi^2}{3\alpha}$ & 1 & $-1$ & $-1$ \\ \hline
	\end{tabular}
	\label{tab:physical_summary}
\end{table*}

\begin{figure*}[ht!]
	\includegraphics[width=0.45\textwidth]{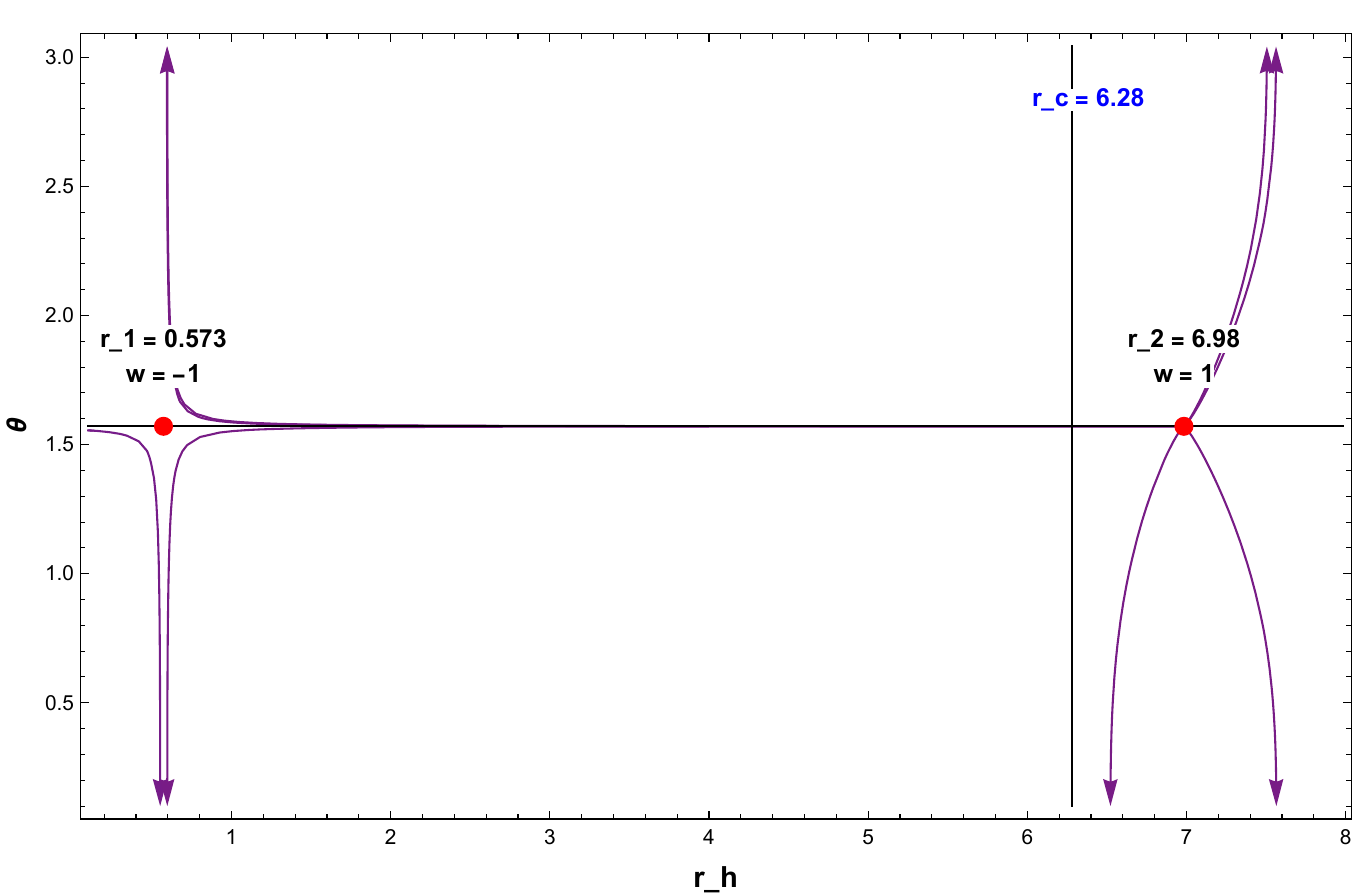}~~~
	\includegraphics[width=0.45\textwidth]{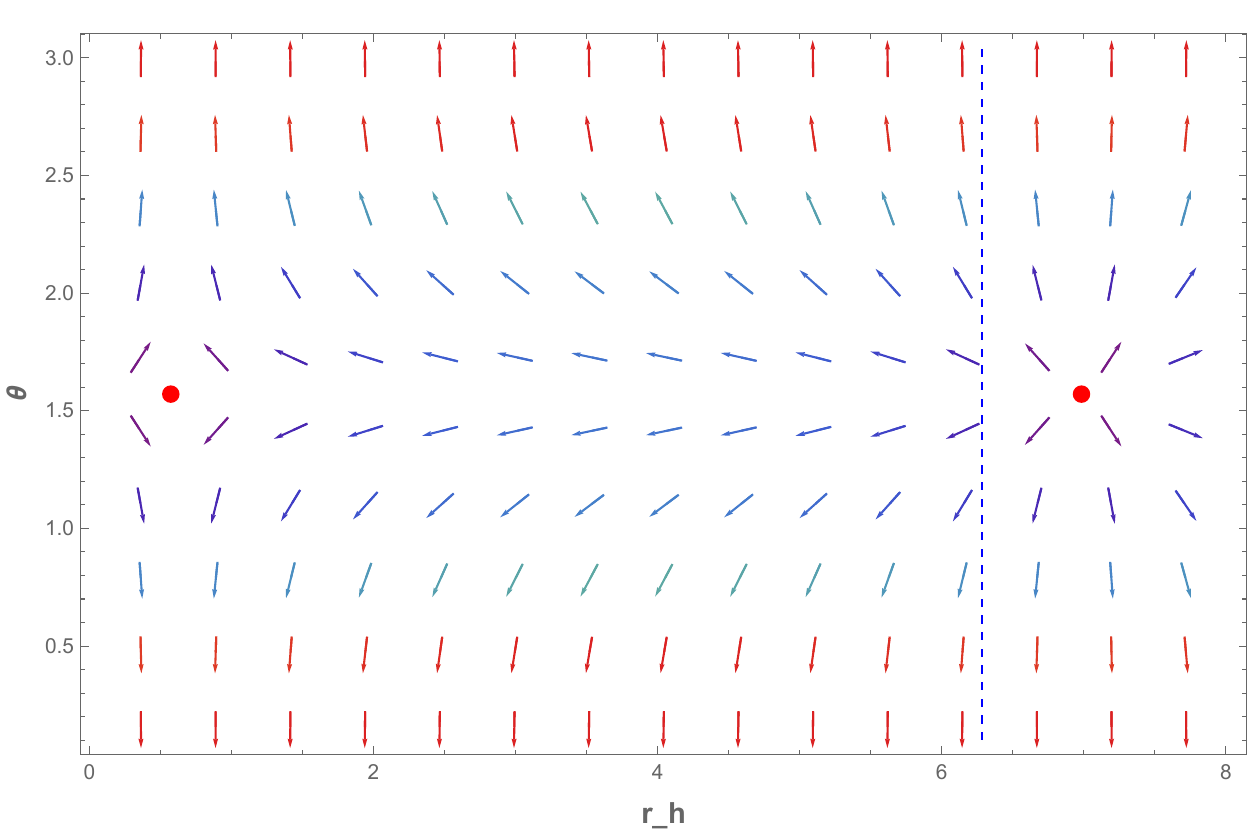}
	\caption{The $r_h$-$\theta$ diagram for $\eta > 0$ (with $\kappa = 1$, 
		$\eta = 0.5$). Two mathematical equilibrium points appear: a saddle at 
		$r_1^*$ and an unstable node at $r_2^*$. However, only the saddle point at 
		$r_1^*$ satisfies the physical constraints $S'(r_h) > 0$ and $T > 0$ 
		(positive ADM mass and positive temperature). The unstable node at $r_2^*$ 
		corresponds to a region where $S'(r_h) < 0$ (negative mass) and $T < 0$ 
		(negative temperature), rendering it unphysical. The physical branch is 
		therefore a single saddle with winding number $w = -1$. The apparent 
		$w = +1$ branch is a mathematical artifact with no physical significance.}
	\label{fig:eta-p}
\end{figure*}

\begin{figure*}[ht!]
	\includegraphics[width=0.45\textwidth]{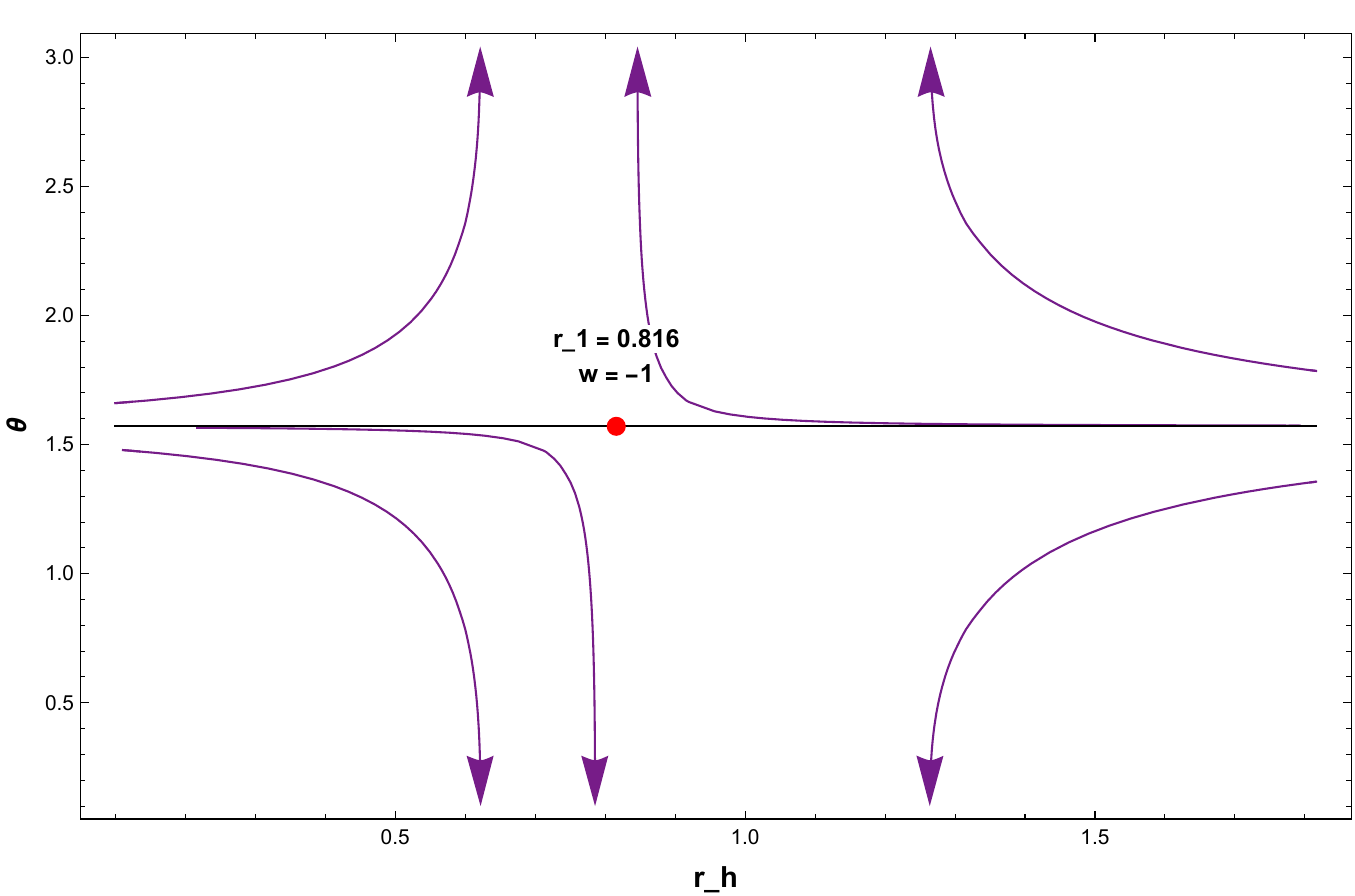}~~~
	\includegraphics[width=0.45\textwidth]{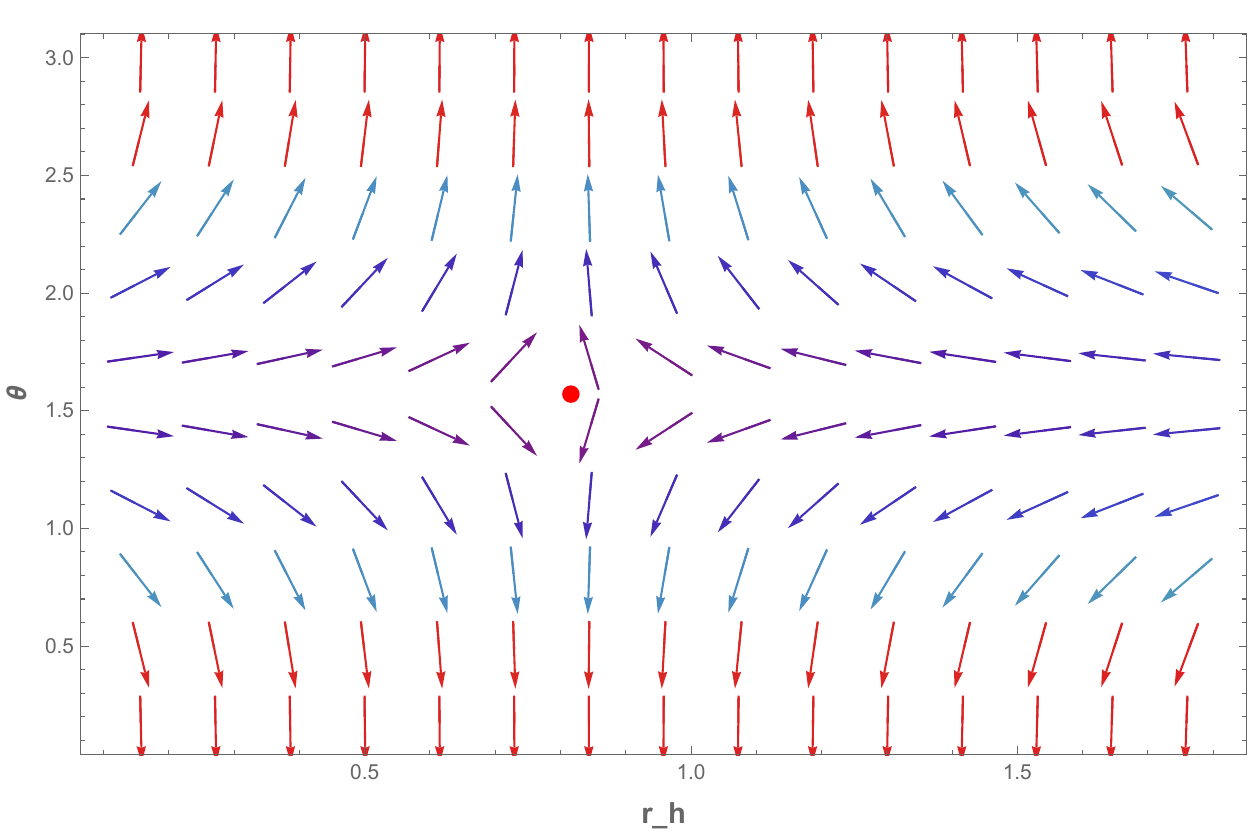}
	\caption{The $r_h$-$\theta$ diagram for $\eta < 0$ (with $\kappa = 1$, $\eta = -0.1$). 
		Here $S'(r_h) > 0$ and $S''(r_h) > 0$ for all $r_h > 0$, so the unique equilibrium point satisfies all physical constraints. 
		The equilibrium is a saddle point with winding number $w = -1$ and total topological charge $W = -1$.}
	\label{fig:eta-n}
\end{figure*}

\section{Thermodynamic (In)stability under Planck-scale Entropy Corrections}\label{sec:vi}

A remarkable feature of the entropy correction \eqref{eq:S_before_redress} with $\eta>0$ is that it mathematically permits two equilibrium branches within the topological formalism. However, when the entropy-geometry correspondence is applied consistently, the ADM mass is given by $M = S'(r_h)/(4\pi)$. For $\alpha > 0$ (i.e., $\eta > 0$), this yields
\begin{equation}
	M = \frac{r_h}{2} - \frac{3\alpha}{4\pi} r_h^2. \label{eq:M_alpha}
\end{equation}
Positivity of the ADM mass requires $S'(r_h) > 0$, which restricts the horizon radius to $r_h < r_c$ where
\begin{equation}
	r_c = \frac{2\pi}{3\alpha} = \frac{\pi\kappa}{\eta}. \label{eq:rc_vi}
\end{equation}
The larger root $r_2^*$ obtained from the equilibrium condition \eqref{eq:quadratic} lies at $r_2^* > r_c$ and therefore corresponds to a negative ADM mass, rendering it unphysical. Consequently, the only physical black hole solutions are those with $r_h < r_c$.

For a Schwarzschild BH ($\alpha = 0$),
\begin{equation}
	S_{\mathrm{BH}} = \pi r_h^2, \qquad M = \frac{r_h}{2},
\end{equation}
the Hawking temperature and specific heat are
\begin{align}
	T &= \frac{1}{4\pi r_h}, \label{eq:T_Schwarzschild} \\
	C &= \frac{dM}{dT} = -2\pi r_h^2 < 0 \quad \text{for all } r_h>0. \label{eq:C_Schwarzschild}
\end{align}
Hence the Schwarzschild solution corresponds to a single thermodynamic branch with negative specific heat; it is always thermodynamically unstable. Its winding number is $w=-1$ and the total topological charge is $W=-1$.

For the cubic entropy correction \eqref{eq:S_before_redress} with $\alpha > 0$, the physical branch ($r_h < r_c$) has $S'(r_h) > 0$, which from Eq.~\eqref{eq:w_physical} gives $w = -1$. The specific heat remains negative, and the black hole is thermodynamically unstable, just as in the Schwarzschild case.
The mathematical second derivative of the 
	generalized free energy at the unphysical root $r_2^* > r_{\max}$ would give 
	$w = +1$, which--if one mechanically applied the standard 
	equivalence--would suggest thermodynamic stability. However, 
	this branch violates the physical constraints: $S'(r_h) < 0$ 
	(implying negative ADM mass) and $T < 0$ (negative Hawking temperature). 
	Therefore, the $w = +1$ classification for this branch is not an 
	indication of a stable black hole phase. Rather, it demonstrates the 
	critical importance of restricting the topological analysis to the physical 
	domain. A reader should not conclude that $w = +1$ always represents a 
	stable black hole; it does so only when $M' > 0$ and $S' > 0$, conditions 
	that the $r_2^*$ branch fails to satisfy.

% The mathematical second derivative of the generalized free energy at the unphysical root $r_2^*$ would be positive, but this root does not correspond to a physical black hole with positive mass.

In general, for the cubic correction at hand, the branch with $w = +1$ corresponds to $S'(r_h) < 0$ and hence negative ADM mass. It is therefore unphysical and must be discarded. It means that the cubic entropy correction at hand yields a total topological charge $W = -1$ for all physical branches, indicating no thermodynamic stabilization. This behavior aligns with the classical Schwarzschild case and with Barrow \cite{Barrow:2020tzx} and R{\'e}nyi \cite{Czinner:2015eyk} entropies, all of which produce only unstable branches. In contrast, logarithmic \cite{Kaul:2000kf,Das:2001ic} and Kaniadakis \cite{Kaniadakis:2002zz} entropies exhibit a distinct topological signature, $W = 0$, arising from the coexistence of stable and unstable branches within their physical regimes. The key difference lies in the functional form: while our cubic correction eventually violates $S'(r_h) > 0$ and $T > 0$ for large $r_h$, logarithmic and Kaniadakis entropies maintain positivity conditions across a wider domain, allowing stable phases to emerge.

On the regime of validity, we emphasize that the entropy correction \eqref{eq:S_before_redress} was derived under the assumption that the Planck-scale correction is small, i.e., $\frac{\eta}{\kappa} r_h \ll 1$ (see the discussion following Eq.~(\ref{eq:ds_intermediate})). The physical regime for $\alpha > 0$ requires $r_h < r_c = \pi\kappa/\eta$, which gives $\frac{\eta}{\kappa} r_h < \pi$. Thus, the upper bound of the physical regime ($r_h \to r_c^-$) corresponds to $\frac{\eta}{\kappa} r_h \to \pi$, where the perturbative expansion parameter is of order unity. At this boundary, the derivation of the entropy correction is no longer under perturbative control. The stable branch (the unphysical $r_2^*$ branch) lies entirely in the region $r_h > r_c$, where $\frac{\eta}{\kappa} r_h > \pi$ and the perturbative expansion is invalid. Therefore, even if one were to ignore the positivity condition $S'(r_h) > 0$, the stable branch would lie outside the domain where the entropy formula itself is reliable.

For completeness, we note that if one extrapolates the entropy formula \eqref{eq:S_before_redress} far beyond its domain of perturbative validity, the model formally predicts a stable branch for $r_h > r_c$. Such an extrapolation, however, lacks theoretical justification and cannot be interpreted as a physical prediction.

The topological classification of the cubic entropy correction is 
summarized in Table~\ref{tab:physical_summary} and illustrated in 
Figs.~\ref{fig:eta-p} and~\ref{fig:eta-n}. For 
$\alpha > 0$ (Fig.~\ref{fig:eta-p}), two mathematical equilibrium 
points exist, but only the saddle at $r_1^*$ is physical; the unstable node 
at $r_2^*$ has $w = +1$ but violates the physical constraints $S'(r_h) > 0$ 
and $T > 0$, and is therefore discarded. For $\alpha < 0$ 
(Fig.~\ref{fig:eta-n}), the unique equilibrium point is a saddle 
with $w = -1$ and lies entirely within the physical domain. In both cases, 
the physical branches yield $W = -1$, identical to the classical 
Schwarzschild case. These results demonstrate that the cubic entropy 
correction arising from Planck-scale modified kinematics does not produce 
thermodynamically stable black holes when physical constraints are properly 
enforced. The apparent $w = +1$ branch is a mathematical artifact that does 
not correspond to a physical black hole state.
\\
For clarity, we summarize the properties of the two mathematical roots in 
Table~\ref{tab:branch_comparison}. The comparison is made for the same 
value of $\tau$ (with $\tau \leq \tau_{\max}$ so that the smaller root is 
physical). The smaller root $r_1^*$ satisfies all physical constraints and corresponds 
to a physical black hole with $w = -1$ (thermodynamically unstable). The 
larger root $r_2^*$ violates $S'(r_h) > 0$ (negative mass), $T > 0$ 
(negative temperature), and lies outside the perturbative regime. Despite 
carrying $w = +1$ mathematically, it does not represent a stable 
black hole.
\begin{table}[h]
	\centering
	\caption{Comparison of the two mathematical branches for $\alpha > 0$. 
		}
	\begin{tabular}{|c|c|c|c|c|c|c|}
		\hline
		\textbf{Branch} & \textbf{Domain} & $\mathbf{S'(r_h)}$ & $\mathbf{M}$ & $\mathbf{T}$ & $\mathbf{w}$ & \textbf{Status} \\
		\hline
		$r_1^*$ & $r_h < r_{\max}$ & $>0$ & $>0$ & $>0$ & $-1$ & \textbf{Physical} (unstable) \\
		\hline
		$r_2^*$ & $r_h > r_{\max}$ & $<0$ & $<0$ & $<0$ & $+1$ & \textbf{Unphysical} (discard) \\
		\hline
	\end{tabular}
		\label{tab:branch_comparison}
\end{table}

\section{Discussion and Conclusion}\label{con}
Determining whether Planck-scale effects can stabilize black holes potentially addresses fundamental questions about black hole evaporation and quantum gravity consistency—namely, the final fate of black holes, the resolution of the information paradox, and the viability of black hole remnants as a physical reality. With this in mind, we have systematically investigated how a phenomenological Modified Dispersion Relation (MDR) of the form \eqref{eq:dsr_dispersion} translates into black hole entropy corrections within the thermodynamic topology formalism. The analysis yields a Planck-scale-modified entropy function \(S = \pi r_h^2 - \alpha r_h^3\) via the Hamilton-Jacobi tunneling method. This cubic deviation from the standard area law serves as a concrete test case for applying the entropy-geometry correspondence of Refs.~\cite{Anand:2025rjg,Anand:2025cer}.

Through the thermodynamic topology approach, we mapped equilibrium states of the black hole to zeros of a vector field \(\boldsymbol{\phi}\) defined on the \((r_h,\theta)\) plane. The winding number associated with each equilibrium point, determined by the convexity of the generalized off-shell free energy \(\mathcal{F}\), provides a classification of thermodynamic stability within the physical domain defined by \(S'(r_h) > 0\) (positive ADM mass) and \(T > 0\) (positive temperature). 

We emphasize that the interpretation of the winding number as a stability 
	indicator depends crucially on the signs of $M'$ and $S'$. Specifically, the 
	standard equivalence $w_i = +1 \iff C > 0$ (stable) holds only when 
	$M' > 0$ and $S' > 0$. If a mathematical equilibrium point lies outside 
	this domain--as does the $r_2^*$ branch for $\alpha > 0$--its winding 
	number cannot be interpreted as indicating thermodynamic stability. The 
	$w = +1$ assigned to the unphysical branch is a purely mathematical 
	classification that does not correspond to a physical black hole state. This 
	is a central point that distinguishes our analysis from a purely algebraic 
	application of the topology method: physical constraints must be imposed 
	before interpreting winding numbers.

We have clarified the relationship between thermodynamic stability and the stability of the auxiliary dynamical system in the \((r_h,\theta)\) plane. While a thermodynamically stable phase would correspond to an unstable node in the auxiliary phase space (see Appendix~\ref{app:stability-relation}), this apparent reversal stems from the sign convention in defining the vector field \(\phi^{r_h} = +\partial\mathcal{F}/\partial r_h\) and does not affect the topological classification. The winding number remains a robust indicator of thermodynamic stability within the physical domain once the sign convention is accounted for.

A consistent application of the entropy-geometry correspondence with all physical constraints leads to the following. For \(\eta < 0\) (\(\alpha < 0\)), the entropy correction is positive, yielding a single unstable equilibrium with \(w = -1\) and \(W = -1\). For \(\eta > 0\) (\(\alpha > 0\)), the correction is negative; while two mathematical roots exist, only the smaller root satisfies \(S'(r_h) > 0\) and \(T > 0\). This physical root gives \(w = -1\) and \(W = -1\), whereas the larger root is unphysical (negative mass or negative temperature).

In both cases, the total topological charge for physical black hole solutions is \(W = -1\), identical to the classical Schwarzschild case. So, our consistent application of the entropy-geometry correspondence demonstrates that no thermodynamically stable branch exists for physical black holes within this framework. More exactly, the stabilization (which would correspond to \(w = +1\)) arises from branch with \(r_2^* > r_c\) which is not health since it lies in a regime where the perturbative expansion used to derive the entropy correction is no longer under control (\(\frac{\eta}{\kappa}r_h \gtrsim \pi\)), further undermining any physical interpretation of this branch. These results are summarized in Tables ~\ref{tab:physical_summary}, and ~\ref{tab:branch_comparison}.

Comparisons with other modified entropies—such as Barrow, Rényi, logarithmic, and Kaniadakis corrections—highlight the importance of imposing physical constraints. While logarithmic and Kaniadakis entropies can yield \(W = 0\) with competing stable and unstable branches within their physical regimes, the cubic correction studied here does not. This difference arises because the would-be stable branch for the cubic entropy violates \(S'(r_h) > 0\) or \(T > 0\), whereas the stable branches of logarithmic and Kaniadakis entropies typically satisfy these conditions.

The thermodynamic topology formalism employed here is not new; it follows directly from Refs.~\cite{Wei:2022dzw,Wei:2024gfz,Anand:2025rjg,Anand:2025cer}. The contribution of this work is threefold. First, we derive the cubic entropy correction from a Planck-scale MDR, thereby providing a specific, physically motivated entropy function to which the formalism can be applied. Second, we identify the importance of imposing physical constraints—namely \(S'(r_h) > 0\) (positive ADM mass) and \(T > 0\) (positive temperature)—when interpreting winding numbers, constraints that are often overlooked in purely algebraic applications of the topology method. Third, we demonstrate that the cubic entropy correction, despite its mathematical interest, does not produce a stable black hole phase when these constraints are properly enforced.

Finally, several limitations of this work should be acknowledged. First, the entropy correction was derived using a perturbative expansion in the small parameter \(\frac{\eta}{\kappa}r_h\); the physical regime for \(\alpha > 0\) extends up to \(r_h \to r_c\) where this parameter is of order unity, placing the derivation at the boundary of perturbative control and requiring a non-perturbative treatment for definitive conclusions near this boundary. Second, the analysis used only a MDR and did not implement the full DSR framework (nonlinear momentum composition, deformed Lorentz transformations); the results should therefore be interpreted as generic to Planck-scale modified kinematics rather than specific to DSR. Third, the entropy-geometry correspondence of Refs.~\cite{Anand:2025rjg,Anand:2025cer} was taken as given; alternative approaches to incorporating entropy corrections into the metric could yield different conclusions.

%Future work should aim to:
%\begin{itemize}
%	\item Derive the entropy correction non-perturbatively from the MDR, without expanding in \(\frac{\eta}{\kappa}r_h\).
%	\item Extend the analysis to rotating or charged backgrounds.
%	\item Investigate whether a fully consistent DSR implementation (including the nonlinear composition law) alters the entropy correction or the stability conclusions.
%	\item Explore other modified entropy functions that might yield physical stable branches while satisfying all constraints.
%\end{itemize}

\appendix
\section{Thermodynamic stability and phase-space dynamics}
\label{app:stability-relation}

In the thermodynamic topology framework, two distinct notions of stability arise: thermodynamic stability of a black-hole equilibrium phase, characterized by the sign of the specific heat \(C\), and dynamical stability in the auxiliary \((r_h,\theta)\) phase space, determined by the eigenvalues of the Jacobian matrix of the vector field \(\boldsymbol{\phi}\). Although these two notions are mathematically connected via the winding number, their physical interpretations differ and, at first glance, appear opposite. This appendix clarifies this relationship.

\subsection{Thermodynamic stability from winding number}

The generalized off-shell free energy is
\begin{equation}
	\mathcal{F}(r_h,\tau) = M(r_h) - \frac{S(r_h)}{\tau}, \label{eq:app_F}
\end{equation}
where \(M(r_h)\) is the black-hole mass expressed as a function of the horizon radius \(r_h\). At equilibrium, \(\tau^{-1} = T(r_h)\), the Hawking temperature. The specific heat at constant volume is
\begin{equation}
C = T \left( \frac{\partial S}{\partial T} \right) = \frac{T}{dT/dS}. \label{eq:app_C}
\end{equation}
As shown in Eqs.~\eqref{eq:specific_heat_formula}–\eqref{c}, within the physical domain where \(M'(r_h) > 0\) and \(S'(r_h) > 0\), the winding number \(w_i\) associated with an equilibrium point \(r_i\) satisfies
\begin{equation}
	w_i = \operatorname{sgn}(C), \label{eq:app_winding_C}
\end{equation}
meaning that
\begin{align}
	w_i &= +1 \;\longleftrightarrow\; C > 0 \quad \text{(thermodynamically stable)}, \label{eq:app_stable} \\
	w_i &= -1 \;\longleftrightarrow\; C < 0 \quad \text{(thermodynamically unstable)}. \label{eq:app_unstable}
\end{align}

\subsection{Phase-space dynamics}

The auxiliary vector field
\begin{equation}
	\boldsymbol{\phi}(r_h,\theta) = \bigl( \phi^{r_h},\; \phi^{\theta} \bigr), 
	\qquad 
	\phi^{r_h} = \frac{\partial\mathcal{F}}{\partial r_h}, \quad
	\phi^{\theta} = -\cot\theta\csc\theta, \label{eq:app_phi}
\end{equation}
defines a fictitious dynamical system on the \((r_h,\theta)\) plane:
\begin{equation}
	\frac{dr_h}{dt} = \phi^{r_h}, \qquad
	\frac{d\theta}{dt} = \phi^{\theta}. \label{eq:app_dynamics}
\end{equation}
Equilibrium points \((r_h^*, \pi/2)\) satisfy \(\phi^{r_h} = 0\) and \(\phi^{\theta} = 0\). The Jacobian at such a point is diagonal:
\begin{equation}
	J = \begin{pmatrix}
		\lambda_r & 0 \\[4pt]
		0 & 1
	\end{pmatrix}, \qquad
	\lambda_r \equiv \left.\frac{\partial\phi^{r_h}}{\partial r_h}\right|_{r_h^*} 
	= \left.\frac{\partial^2\mathcal{F}}{\partial r_h^2}\right|_{r_h^*}. \label{eq:app_Jacobian}
\end{equation}
The eigenvalues are \(\lambda_1 = \lambda_r\) and \(\lambda_2 = 1\).

Throughout 
	this analysis, we have emphasized that the winding number $w_i$ is a 
	topological invariant that, on its own, does not encode physical information 
	about whether a black hole is stable. The physical interpretation of $w_i$ 
	as a stability indicator requires two ingredients: (i) the branch must lie 
	in the physical domain where $M'(r_h) > 0$ and $S'(r_h) > 0$, and (ii) the 
	standard thermodynamic relations must apply. When these conditions hold, 
	$w_i = +1$ corresponds to $C > 0$ (thermodynamically stable) and $w_i = -1$ 
	corresponds to $C < 0$ (thermodynamically unstable). If a branch violates 
	these conditions, its winding number is a purely mathematical quantity 
	without physical significance. In particular, the apparent $w = +1$ branch 
	in our analysis is unphysical because it violates $S'(r_h) > 0$ and 
	$T > 0$, and therefore should not be interpreted as evidence of a stable 
	black hole phase.

Because \(\lambda_2 = 1 > 0\), the equilibrium is always unstable along the \(\theta\)-direction. The stability in the \(r_h\)-direction is governed by \(\lambda_r\):

\(\lambda_r < 0\): eigenvalues \((-, +)\) → \textbf{saddle} in the \((r_h,\theta)\) plane.

 \(\lambda_r > 0\): eigenvalues \((+, +)\) → \textbf{unstable node} in the \((r_h,\theta)\) plane.

Since \(w_i = \operatorname{sgn}(\lambda_r)\) from Eq.~\eqref{eq:winding_second_deriv}, we have:
\begin{align}
	w_i &= +1 \;\longleftrightarrow\; \lambda_r > 0 \;\longleftrightarrow\; \text{unstable node in } r_h\text{-direction}, \label{eq:app_wplus} \\
	w_i &= -1 \;\longleftrightarrow\; \lambda_r < 0 \;\longleftrightarrow\; \text{saddle in } r_h\text{-direction}. \label{eq:app_wminus}
\end{align}
The connection between \(w_i = +1\) (thermodynamically stable) and \(\lambda_r > 0\) (unstable in the auxiliary phase space) may seem misleading at first, but the resolution lies in the distinct roles of the two dynamics. Thermodynamic stability concerns the response of the black hole to thermal fluctuations under physical time evolution: a positive specific heat means that if the black hole absorbs a small amount of heat, its temperature increases, which in turn enhances Hawking radiation and tends to restore the original state—a stabilizing feedback. In contrast, phase-space stability refers to the behavior of the fictitious gradient-flow system \(\dot{r}_h = +\partial\mathcal{F}/\partial r_h\), which is constructed solely to locate equilibria (\(\partial\mathcal{F}/\partial r_h = 0\)) and assign a topological winding number; it does not describe real thermodynamic evolution.

In the gradient-flow picture, a thermodynamically stable equilibrium (\(C > 0\)) corresponds to a minimum of \(\mathcal{F}\) with respect to \(r_h\) for fixed \(\tau\). However, because our flow is \(\dot{r}_h = +\partial\mathcal{F}/\partial r_h\) (without a minus sign), a minimum of \(\mathcal{F}\) is a \emph{repeller} in the \(r_h\)-direction: small perturbations grow away from the equilibrium. If we had chosen \(\dot{r}_h = -\partial\mathcal{F}/\partial r_h\) (the usual gradient descent), a minimum would become an attractor, yet the winding-number classification would remain unchanged—the zeros of \(\boldsymbol{\phi}\) and the sign of \(\partial^2\mathcal{F}/\partial r_h^2\) are unaffected by an overall sign in \(\phi^{r_h}\). Thus, the winding number directly reflects thermodynamic stability (\(w = +1\) for \(C > 0\), \(w = -1\) for \(C < 0\)) within the physical domain (\(M'(r_h) > 0\), \(S'(r_h) > 0\)). The auxiliary dynamical system is merely a computational tool; its stability properties should not be confused with physical thermodynamic stability, and the apparent sign reversal is simply a matter of convention.

\bibliographystyle{apalike}

\end{document}